\documentclass[aps,twocolumn,floatfix,altaffilletter,superscriptaddress,preprintnumbers,
               tightenlines,showpacs,showkeys,notitlepage,nofootinbib]{revtex4-2}

\usepackage{amsmath,amssymb} 
\usepackage{graphicx} 
\usepackage[export]{adjustbox} 
\usepackage{url} 
\usepackage[dvipsnames]{xcolor} 
\usepackage{siunitx} 
\usepackage{etoolbox,orcidlink} 
\usepackage{tabularx,booktabs} 
\usepackage{microtype}
\usepackage[normalem]{ulem} 
\usepackage{braket}
\usepackage{comment}

\usepackage{titlesec}
\usepackage{hyperref}

\newcommand{\f}{\frac}

\renewcommand{\d}{\hspace{0.2mm}\mathrm{d}}
\newcommand{\field}{\varphi}
\newcommand\MSbar{$\overline{\rm MS}$}

\titleformat{\section}[runin]
  {\normalfont\normalsize\bfseries}
  {}
  {0pt}
  {}

\titlespacing*{\section}
  {\parindent}
  {1.5ex}
  {\fontdimen2\font}

\makeatletter
\renewcommand*{\toclevel@section}{2}
\makeatother

\makeatletter
\newcommand{\myendmatterrule}{%
  \par
  \onecolumngrid@push
  \begingroup
    \let\phantomsection\relax
    \let\addcontentsline\@gobblethree
    \let\hyper@anchorstart\@gobble
    \let\hyper@anchorend\relax
    \baselineskip26\p@
    \bib@device{\textwidth}{245.5\p@}%
  \endgroup
  \nobreak\@nobreaktrue
  \addvspace{19\p@}%
  \par
  \onecolumngrid@pop
}
\makeatother

\makeatletter
\def\thesection{\arabic{section}}%
\def\p@section{}%
\def\thesubsection{\thesection.\arabic{subsection}}%
\def\p@subsection{}%
\def\thesubsubsection{\thesubsection.\arabic{subsubsection}}%
\def\p@subsubsection{}%
\def\appendix{%
    \par
    \setcounter{section}\z@
    \setcounter{subsection}\z@
    \setcounter{subsubsection}\z@
    \def\thesubsection{\thesection.\arabic{subsection}}%
    \def\thesubsubsection{\thesubsection.\arabic{subsubsection}}%
    \def\p@subsection{}%
    \def\p@subsubsection{}%
    \@addtoreset{equation}{section}%
    \def\theequation@prefix{\thesection}%
    \addtocontents{toc}{\protect\appendix}%
    \@ifstar{%
    \def\thesection{\unskip}%
    \def\theequation@prefix{A.}%
    }{%
    \def\thesection{\Alph{section}}%
    }%
}%
\makeatother

\begin{document}

\title{Tachyonic particle production in strongly supercooled phase transitions}

\newcommand{\UWaffiliation}{%
  Faculty of Physics, University of Warsaw,
  ul. Pasteura 5, 02-093 Warsaw, Poland%
}

\author{Mateusz Kulejewski \orcidlink{0009-0004-5269-7003}}
\email{m.kulejewski@uw.edu.pl}
\affiliation{\UWaffiliation}

\author{Bogumi{\l}a {\'S}wie{\.z}ewska \orcidlink{0000-0003-0169-211X}}
\email{bogumila.swiezewska@fuw.edu.pl}
\affiliation{\UWaffiliation}

\author{Jorinde van de Vis \orcidlink{0000-0002-8110-1983}}
\email{jorinde.van.de.vis@cern.ch}
\affiliation{Theoretical Physics Department, CERN,
             1211 Geneva 23, Switzerland}
\affiliation{Department of Physics and Helsinki Institute of Physics, PL 64, FI-00014, University of Helsinki, Finland
             }

\date{\today}

\preprint{CERN-TH-2026-231, HIP-2026-14/TH} 

\begin{abstract}
\noindent
In strongly supercooled cosmological phase transitions, which generically lead to strong, observable gravitational-wave backgrounds, the scalar field typically transitions to a value well below the minimum of its potential.
We study particle production caused by the subsequent evolution of the field.
Because some modes feature temporarily tachyonic masses, they can be efficiently produced, and a significant fraction of the energy density released in the phase transition can 
get converted to particles rather than gradient energy. 
This has consequences for the shape and amplitude of the gravitational-wave spectrum caused by the interactions in bubbles
and sources a new, high-frequency signal.
\end{abstract}

\maketitle

\pdfbookmark[1]{Contents}{bookmark:contents}

Supercooled first-order phase transitions (PTs), taking place at temperatures much below the critical temperature, typically proceed slowly, feature a large latent-heat release and consequently generically yield very strong, observable gravitational-wave (GW) signals~\cite{Randall:2006py, Konstandin:2011dr, Kubo:2016kpb, Jaeckel:2016jlh}.
They are therefore a prime target for GW searches with LISA~\cite{Caprini:2019egz, Ellis:2020nnr, 
Kierkla:2023von, Athron:2023xlk, Caprini:2024hue,Kierkla:2025qyz} 
and have been studied in the context of the recent PTA data~\cite{NANOGrav:2023gor, EPTA:2023fyk, Reardon:2023gzh, Xu:2023wog, NANOGrav:2023hvm, Figueroa:2023zhu, Ellis:2023oxs, Balan:2025uke, Bringmann:2026xcx,  Biondini:2026uds}. 
Moreover, supercooled PTs have attracted considerable interest as an environment for primordial black hole (PBH) production~\cite{Hashino:2021qoq, Gouttenoire:2023naa, Lewicki:2023ioy, Franciolini:2025ztf, Kierkla:2025vwp, Ning:2026nfs}.

Supercooling  naturally occurs in models with strong dynamics or extra dimensions~\cite{Randall:2006py,
Konstandin:2010cd,
Konstandin:2011dr,
vonHarling:2017yew,
Bruggisser:2018mrt,
Baldes:2021aph},
and with perturbative classically scale-invariant potentials~\cite{Hambye:2013dgv,Kubo:2016kpb,Jaeckel:2016jlh,Hashino:2016rvx,Jinno:2016knw,Marzola:2017jzl}, which are the focus of this work.
There, the lack of an explicit mass term in the potential makes the thermally induced barrier last until very low temperatures at field values that are much below the scale of the radiatively generated minimum. 
Consequently, the scalar field transitions to a field value much below the minimum~\cite{Kierkla:2022odc} and nucleation is followed by a period of scalar field rolling and oscillations, partially through a region of negative effective mass. Studying this evolution and the associated particle production is the subject of this work.

In vacuum transitions, all of the released energy gets transferred to the bubble walls, resulting in a distinct GW spectrum \cite{Kosowsky:1992vn, Cutting:2018tjt, Cutting:2020nla}.
Similarly, in the most supercooled phase transitions, where the plasma is severely diluted by a phase of thermal inflation, the GW spectrum is expected to be sourced by bubble-wall collisions~\cite{Ellis:2019oqb}.
We challenge this picture by studying particle production at the early stage of the evolution of the bubble.
The scalar field modes have a tachyonic mass during the rolling phase, leading to copious particle production via tachyonic resonance.
We find that a significant fraction of the released vacuum energy can be converted into fluctuations, rather than the gradient energy of the background field.
This particle production, not considered in the literature before, can have significant phenomenological consequences: 
change in the spectral shape and amplitude of the predicted GW signal;
modified dynamics of the bubble walls during the expansion phase; 
an additional GW signal 
sourced at a much higher energy, associated with the resonant particle production;
modified reheating history.

Earlier works have focused on particle production from bubble nucleation, expansion, and collisions
\cite{Rubakov:1984pa, Zhang:2010qg, Braden:2014cra, Shakya:2023kjf, Mansour:2023fwj, Inomata:2024rkt, GarciaGarcia:2026bvd, An:2026sdu, Ghoshal:2026pew, Sengupta:2025cdm, Sengupta:2025jah, Sengupta:2026xlp}.
Production from transition radiation was studied in~\cite{Azatov:2020ufh, Azatov:2023xem, Jinno:2026gsh}, with the internal structure of a thick wall accounted for in~\cite{Jinno:2026gsh}.
Tachyonic production during a homogeneous strongly supercooled phase transition was analysed in~\cite{Schmitt:2024pby}, 
and reheating at the final stage of a PT, after bubble collisions in~
\cite{Mansour:2026sdx, Rescigno:2025ong}.
While the formalism for particle production 
sourced by the background evolution during a first-order PT, beyond the thin-wall approximation, was already developed in~
\cite{Yamamoto:1994te, Hamazaki:1995dy}, 
in this work, we will apply it to the case of tachyonic particle production for the first time.
Since particle production is very efficient in our case, we modify the formalism to include backreaction in the equation of motion (EoM) of the background, such that energy is conserved \cite{Boyanovsky:1994me, Boyanovsky:1996sq, Kofman:1997yn,Herring:2024nqg}.
Mode-mode scattering is, however, not included in our linearized mode equations and would require a lattice simulation, the state-of-the-art for cosmological particle production \cite{Felder:2000hq,Child:2013ria, Nguyen:2019kbm, vandeVis:2020qcp,Figueroa:2021yhd, Masubuchi:2026eau}.
Due to the growth of the bubble, a large range of momenta gets excited, making the problem difficult to track on a lattice. Therefore, our current semi-analytical
study is an essential step forward, 
which can also be used as a benchmark for future non-linear studies.

\section{Model}
We consider the SU(2)cSM, a model well-studied in the literature~\cite{Hambye:2013dgv, Carone:2013wla, Khoze:2014xha, Pelaggi:2014wba, Karam:2015jta, Khoze:2016zfi, Chataignier:2018kay, Hambye:2018qjv, Baldes:2018emh, Prokopec:2018tnq, Marfatia:2020bcs, Kierkla:2022odc, Kierkla:2023von, Kierkla:2025qyz}, 
consisting of the conformal SM (without an explicit Higgs mass term), extended with a dark SU(2) sector, with gauge coupling $g_X$ and a scalar doublet of the new SU(2), which is a singlet of the SM gauge group, coupled to the Higgs field via a portal. This model is a concrete working example, however, we expect similar effects in any model featuring radiative symmetry breaking.
The effective potential can be expressed as a function of the background fields of the radial components of both scalar fields.
The phase transition proceeds along the direction of the new scalar field, $\varphi$~\cite{Prokopec:2018tnq, Kierkla:2022odc}, 
thus for nucleation, the Higgs direction can be ignored.
However, the subsequent rolling phase proceeds in the Higgs and the $\field$-direction. 
Nonetheless, for simplicity, we only consider the effective potential and the dynamics for the background field in the radial direction of the new doublet.

The explicit formulas for the effective potential can be found in the End Matter. 
In the zero-temperature potential, we include the tree-level $\lambda_{\field}\field^4$ term and the one-loop Coleman--Weinberg correction from the gauge bosons.
The scalar-field contribution is not included because it is subleading and, more importantly, in the following we will consider the production of the scalar-field fluctuations, so we cannot resum them into the effective potential.
Furthermore, we include thermal corrections to the effective potential from the dark vector bosons. 
As we consider vacuum transitions (see discussion in the next section), we will not implement the recently developed treatment of thermal resummations for nucleation in supercooled transitions based on dimensional reduction~\cite{Kierkla:2023von, Kierkla:2025qyz},
but rather implement the simpler Arnold-Espinosa~\cite{Arnold:1992rz} daisy-resummation.
Since the field excursion from the thermal barrier to the radiatively generated minimum is large,  it is crucial to use the renormalisation-group (RG)-improved potential. We follow the procedure of~\cite{Kierkla:2022odc, Kierkla:2023von}, using the running couplings in the potential and taking $\mu= \textrm{max}\left(M_X(\field),\ \pi T \right)$.

\section{Bubble nucleation} 
To determine the particle production after nucleation, we will consider a \emph{vacuum} transition.
Realistically, in the SU(2)cSM the Euclidean action of an O(3)-symmetric thermal bounce is always lower than that of the O(4) vacuum one, making thermal nucleation the dominant process. 
Nonetheless, we argue that O(4)-symmetric vacuum transitions and the subsequent SO(1,3)-symmetric 
evolution of the field are a good playground for testing the hypothesis of particle production during early stages of bubble expansion.
First, at a given temperature, a bubble associated with a vacuum transition is always thinner than a corresponding thermal one.\footnote{This is due to the symmetry-related friction term in the bounce equation. E.g.~in eq.~\eqref{eq:bounce} for the O(3)-symmetric case the second term would read $\f{2}{\xi}\field'(\xi)$. 
This would correspond to evolution with lower friction, and thus a lower escape point (thicker bubble) than in the O(4)-symmetric case.} 
Therefore, considering vacuum transitions yields a conservative estimate of particle production.
In addition, with this assumption, the dynamics of the scalar background depend on one parameter only \cite{Yamamoto:1994te},
and the different mode equations decouple.
Moreover, supercooled PTs proceed after a period of thermal inflation; thus, the primordial plasma is severely diluted, and we do not expect it to affect the post-nucleation dynamics of the scalar field significantly.

The profile of the nucleated bubble is given by the bounce configuration. 
In $D=4$ Euclidean spacetime ($\tau = it$), this yields the following $O(4)$-symmetric equation
\begin{equation}
\label{eq:bounce}
    \f{\d^2\varphi}{\d\xi^2} \, + \, \frac{3}{\xi} \, \f{\d\varphi}{\d\xi} \, - \, V'(\varphi) = 0,
\end{equation}
where $\xi^2 \equiv \tau^2 + r^2$ is the Euclidean radial coordinate,
and the prime denotes the derivative w.r.t.\ $\varphi$. 
The boundary conditions read $\varphi(\xi\to\infty)=0$ and $\f{\d\varphi}{\d\xi}(\xi=0)=0$. 

The bounce profile interpolates between the false vacuum outside the bubble and a nonzero field value inside the bubble, $\varphi_0$ (the escape point). In strongly supercooled phase transitions, typically $\varphi_0/w \sim O(10^{-3})$, with $w$ the field value at the minimum.

\section{Evolution of the field}
Analytic continuation of the bounce solution to real time results in an SO(1,3)-symmetric solution~\cite{Coleman:1977py}. 
Therefore, 
we decompose the full quantum field into the SO(1,3)-symmetric background and quantum fluctuations.\footnote{In principle, the full complex field should be decomposed into modes and the Higgs modes should be included as well. Here, for simplicity, we focus only on the real component of the field. } This can be done naturally in Milne coordinates $(s,\chi,\Omega)$ as
\begin{equation}
    \hat{\varphi}(s,\chi,\Omega) = \varphi(s) + \delta\hat\varphi(s,\chi,\Omega),
\end{equation}
where the Milne coordinates are defined via
$t = s \cosh\chi$ and $r = s\sinh\chi$,
and $\Omega$ are the coordinates on the two-sphere. Then the EoMs for $\varphi(s)$ and $\delta\hat\varphi(s,\chi,\Omega)$  up to $O(\hbar)$ are \cite{Herring:2024nqg}
\begin{align}
    \ddot\varphi \, + \, \frac{3}{s} \, \dot\varphi \, + \, V'(\varphi) \, + \, \frac{1}{2}V'''(\varphi) \braket{\delta\hat\varphi^2} & = 0, \label{eq:EOM_background_no_decomposition} \\
    \delta\ddot{\hat\varphi} \,+\, \frac{3}{s}\, \delta\dot{\hat\varphi} \,-\, \frac{1}{s^2} \, \nabla^2_{\mathbb{H}^3} (\delta\hat\varphi) \,+\,  V''(\varphi) \, \delta\hat\varphi &= 0, \label{eq:EOM_fluctuations_no_decomposition}
\end{align}
where the dot signifies the derivative w.r.t.\ $s$, $\nabla^2_{\mathbb{H}^3}$ is the Laplacian on the three-dimensional unit hyperboloid, and we neglect the Hubble friction as it is suppressed w.r.t.\ the $3/s$ friction term, with $H \sim O(10^{-13}) \text{ GeV}$ while we consider $s \sim O(1-10) \text{ GeV}^{-1}$.
The fluctuations described by Eq.~\eqref{eq:EOM_fluctuations_no_decomposition} can be decomposed on the three-dimensional hyperboloid as
\begin{align}
    \delta\hat\varphi(s,\chi,\Omega) = \int\limits_0^\infty \d p \sum\limits_{l=0}^{\infty} & \sum\limits_{m=-l}^{l} \Big[  \delta\varphi_{plm}(s) \, Y_{plm}(\chi,\Omega) \, \hat{b}_{plm} \nonumber \\ 
    & + \delta\varphi_{plm}^*(s) \, Y_{plm}^*(\chi,\Omega) \, \hat b^\dagger_{plm} \Big],
    \label{eq:mode_decomposition}
\end{align}
where $\hat b_{plm}, \hat b^\dagger_{plm}$ are the Milne annihilation and creation operators satisfying $[\hat b_{plm}, \hat b^\dagger_{p'l'm'}] = \delta_{ll'}\delta_{mm'}\delta(p-p')$ and $Y_{plm}$'s are the eigenfunctions of the three-dimensional hyperbolic Laplacian; see also the End Matter. 

Using this decomposition, the EoMs for the fluctuations can be rewritten as a set of equations for every $plm$-mode, 
\begin{align}
    \delta\ddot\varphi_{p} \,+\, \frac{3}{s}\, \delta\dot\varphi_{p} \,+\, \left(\frac{p^2+1}{s^2} \, \,+\,  V''(\varphi)\right) \, \delta\varphi_{p} & = 0,
    \label{eq:EOM_fluctuations}
\end{align}
where we suppressed the $lm$ indices as the equations depend only on  $p$. Eqs.~\eqref{eq:EOM_background_no_decomposition}, with $\braket{\delta\hat \varphi^2}$ determined from the $\delta \varphi_p$, and \eqref{eq:EOM_fluctuations} form a closed set of equations.
For the details of the renormalisation of the backreaction term in the background's EoM \eqref{eq:EOM_background_no_decomposition}, we refer to the End Matter. 
The initial condition for eq.~\eqref{eq:EOM_background_no_decomposition} is given by the escape point of the bounce solution, $\varphi(s\to 0)=\varphi_0$. 
Following \cite{Yamamoto:1994te, Hamazaki:1995dy}, we find the initial conditions for the $\delta\varphi_p$ -- the seed fluctuations coming from the nucleation of a bubble --  by
solving the mode equations in the Euclidean region,
\begin{equation}
    \f{\d^2(\delta\varphi_p)}{\d \xi^2} + \f{3}{\xi}\f{\d(\delta\varphi_p)}{\d\xi} + \left( \f{p^2+1}{\xi^2} - V''(\varphi) \right) \delta\varphi_p = 0,
    \label{eq:bounce_modes}
\end{equation}
with the initial state given by the metastable, Minkowski vacuum,
$\delta\varphi_p(\xi\to\infty) = K_{ip}(\mu\xi)/\xi$,
where $\mu^2 \equiv V''(\varphi(\xi\to\infty))= V''(0)$ is the mass in the metastable vacuum and $K_\nu(z)$ is the modified Bessel function of the second kind.
We have checked that in the Euclidean region, where there is no significant particle production, the backreaction term of Eq.~\eqref{eq:EOM_background_no_decomposition} does not influence the evolution significantly.
The initial conditions for the modes are then the analytic continuation of the solution of Eq.~\eqref{eq:bounce_modes} to real time at $s=i\xi\to0$.

\section{Energy density and pressure}
We calculate the energy density, $\rho$, and radial pressure of the fluctuations, $P_r$, for a Minkowski observer by a transformation of the stress-energy tensor from Milne to Minkowski coordinates,
\begin{align}
    \rho & = \tilde{\rho}\,\cosh^2\chi  + \tilde{P}\,\sinh^2\chi \, ,
    \label{eq:energydens}
    \\
    P_r&=\tilde{\rho}\,\sinh^2\chi + \tilde{P}\,\cosh^2 \chi\, ,
    \label{eq:press}
\end{align}
where the Milne fluctuations' energy density and isotropic pressure read
\begin{align}
    \tilde{\rho} & = \int\limits_0^\infty \f{p^2\d p}{4\pi^2}\Big[|\delta\dot\varphi_p|^2 + \f{p^2+1}{s^2}|\delta\varphi_p|^2 + m^2|\delta\varphi_p|^2  \Big],
    \label{eq:milne_energy_density}
    \\
     \tilde{P}& = \int\limits_0^\infty \f{p^2\d p}{4\pi^2}\Big[|\delta\dot\varphi_p|^2 - \f{p^2+1}{3s^2}|\delta\varphi_p|^2 - m^2|\delta\varphi_p|^2  \Big],
    \label{eq:milne_pressure}
\end{align}
where $m^2 = V''(\varphi(s))$ is the field-dependent mass.
These quantities, as written, are divergent and need to be renormalized, for which we again refer to the End Matter.
The Minkowski isotropic pressure, relevant for the computation of the equation of state (EoS), is obtained by averaging over the three spatial directions, $P  = \frac{1}{3}\left(P_r+2P_\perp\right)$,
where the transverse pressure $P_\perp=\tilde{P}$, since the transformation from Milne to Minkowski coordinates affects only the radial direction and therefore leaves the transverse components of the stress-energy tensor unchanged.

\section{Results}
We solve the coupled Eqs.~\eqref{eq:EOM_background_no_decomposition} and \eqref{eq:EOM_fluctuations} in {\tt Python} for a discretised momentum lattice $p_i$
for the benchmark (BM) points listed in Table~\ref{tab:benchmark}.
By using the O(4) rather than the O(3) bounce, we drop the realistic nucleation criterion, and thus the temperature $T$ becomes an input parameter, which can be freely chosen.
The considered temperatures are all above the nucleation temperature determined in \cite{Kierkla:2022odc}, such that our estimate of particle production is conservative.
Additional details on the numerical implementation are provided in the End Matter.

\begin{table}[ht]
    \centering
    \begin{tabular}{|c|c|c|c|c|}
        \hline
        $g_X$ & $M_X$ [GeV] & $T$ [GeV] & $T_n$ [GeV] \cite{Kierkla:2022odc} & BM nr\\ 
        \hline
        0.8 & $3\times 10^3$ & 9, 10, 15 & 0.44 & 1a, 1b, 1c \\
        0.9 & $3\times 10^3$ & 15 & 0.80 & 2 \\ 
        0.8 & $10^4$ & 30 & 1.39 & 3\\
        \hline
    \end{tabular}
    \caption{Benchmark points.}
     \label{tab:benchmark}
\end{table} 

\begin{figure}[tb]
    \centering
    \includegraphics[width=\linewidth]{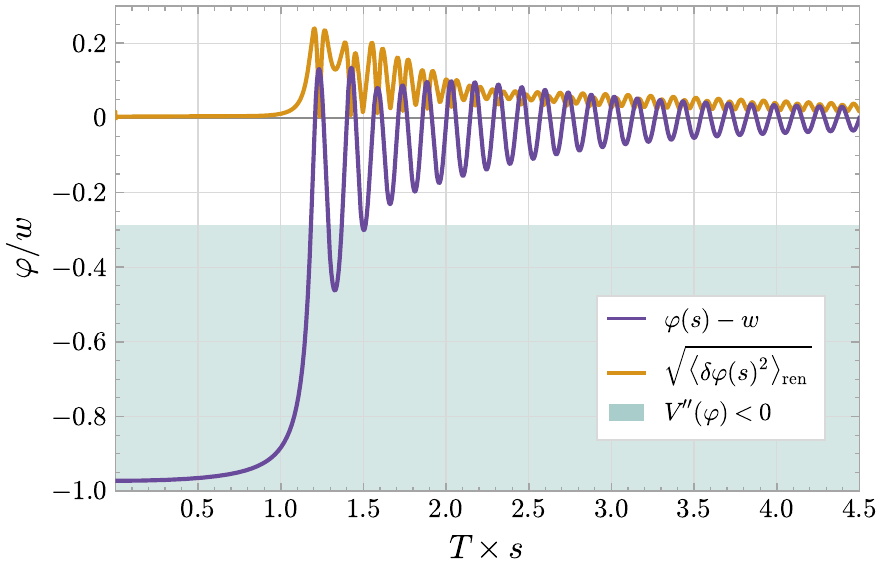}
    \caption{Evolution of the background field $\varphi(s)$ and the RMS of the (renormalized) fluctuations $\sqrt{\braket{\delta\varphi(s)^2}_{\text{ren}}}$ for BM 1a in Milne time rescaled by temperature $T \times s$.
    }
    \label{fig:evolution}
\end{figure}

We show the evolution of the field and the root mean square (RMS) of the (renormalized) fluctuations in Fig.~\ref{fig:evolution}.
The fluctuations start getting produced copiously as soon as the background field starts rolling down.
The impact of the backreaction of produced fluctuations is clearly visible: the strong production during the rolling phase and the first oscillation dampens the amplitude of\ $\field$. 

\begin{figure}[tb]
    \centering
    \includegraphics[width=\linewidth]{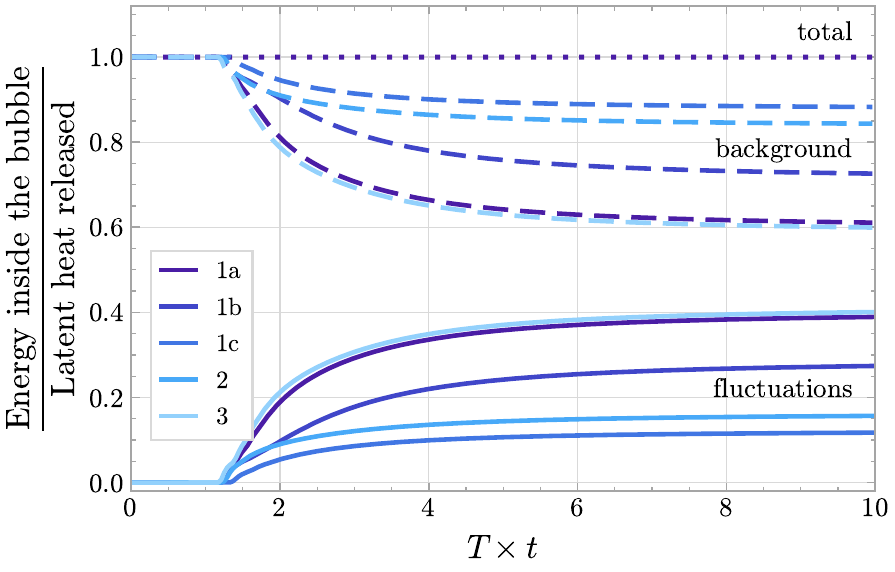}
    \caption{Integrated energy inside a bubble at time rescaled by temperature $T\times t$, normalized with respect to the energy in a corresponding volume of false vacuum
    for all BMs from Table~\ref{tab:benchmark}. The different BMs are shown with shading, while the contributions to the energy budget are shown in solid, dashed, and dotted lines for the fluctuations, background, and total, respectively. The total is calculated as a conservation check and is shown only for BM 1a, as all others overlap.  
    }
    \label{fig:energy_parameters}
\end{figure}

In Fig.~\ref{fig:energy_parameters}, we present the evolution of the bubble's energy budget for all BM points. 
It shows that a significant fraction of the released energy budget can go into the fluctuations of the field. 
Importantly, we can see that the produced fluctuations continue to influence the overall energy budget for later times.
In Milne coordinates, the production visible in Fig.~\ref{fig:evolution} happens at small values of $s$, but in Minkowski coordinates, this corresponds to a continuous production,  localised in the vicinity of the bubble wall.
Moreover, the amount of energy transferred to the fluctuations increases with decreasing temperature. It can be intuitively understood that for lower temperatures, the bounce solution is thicker and starts from a smaller value, which allows the field to stay in the slow-rolling phase longer. This means that once the background field rolls towards the global minimum, the oscillations are damped by a smaller $3/s$ factor and the associated particle production effect is stronger.

\begin{figure}[tb]
    \centering
    \includegraphics[width=\linewidth]{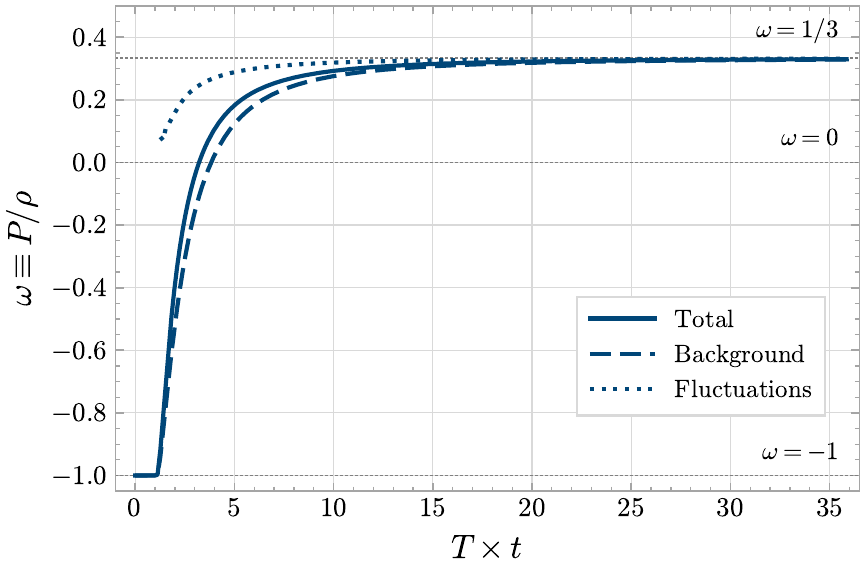}
    \caption{Equation of state parameter averaged over the whole bubble at time rescaled by temperature $T \times t$ for BM 1a. The total EoS parameter is shown in solid, while the corresponding EoS parameters for the background and fluctuations are shown in dashed and dotted lines, respectively.
    }
    \label{fig:eos}
\end{figure}

The EoS parameter, $\omega \equiv P/\rho$,  averaged over the whole bubble, with contributions from both background and fluctuations, is shown in Fig.~\ref{fig:eos}. We can see that all the contributions to the EoS parameter quickly approach the relativistic $\omega=1/3$. Although the particles need not be relativistic at the time of production, they follow the bubble wall and quickly accelerate to relativistic momenta.

\begin{figure}[tb]
    \centering
    \includegraphics[width=\linewidth]{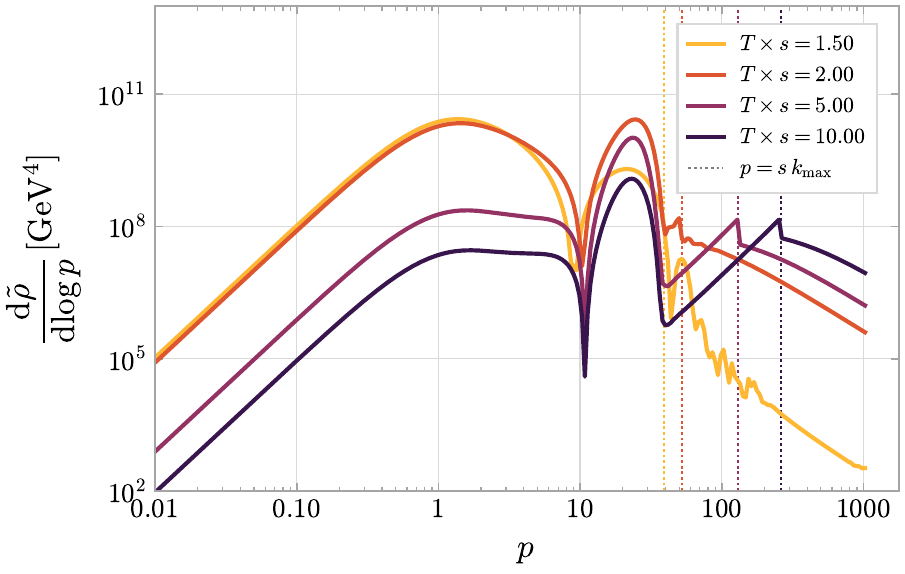}
    \caption{Milne energy density spectrum of fluctuations at the bubble center at given Milne times for BM 1a. The evolution in time is shown with shading, with darker lines corresponding to later times. The consecutive tachyonic vacuum subtraction cutoff lines are shown as dotted vertical lines. 
        }
    \label{fig:spectrum}
\end{figure}

In Fig.~\ref{fig:spectrum}, we plot the spectrum in $p$ of the Milne energy density of produced fluctuations at the bubble center at consecutive Milne times during bubble evolution. The smallest time $T\times s = 1.5$ is chosen to avoid the tachyonic regime, in which the particle momenta and correspondingly the energy spectrum  are ill-defined. We point out the band of highly produced modes around $p \sim 20$. 
As the energy density of fluctuations decreases with $s$, the bubble energy budget is dominated by the contribution from the low-$s$ (close to the light cone) region.
Moreover, we note that renormalization with possible tachyonic instability involves an additional lower momentum cutoff $\Theta(p/s-k_{\text{max}})$, with $k_{\text{max}}=\max(\sqrt{-V''(\varphi)})$, see also \cite{Herring:2024nqg}. The impact of this cutoff is visible in the plot, with the divergence growing until the lower cutoff.
In principle, this divergence could be removed by a better-suited renormalization scheme. However, it becomes important for the total energy budget only at late Milne times, when the contribution to the Minkowski energy is anyway negligible compared to the earlier contributions.

\section{Discussion}
We have demonstrated that models with strong supercooling likely feature efficient particle production starting immediately after the formation of bubbles. 
An $\mathcal O(1)$ fraction of the energy gets converted to particles, and after the initial increase, this fraction remains constant over time.
Here, we have focused on the production of the scalar radial mode itself, but we expect efficient production of all degrees of freedom that develop a tachyonic mass.
Our findings can have far-reaching consequences for the GW spectrum of the most supercooled PTs, which is typically very strong, but follows from the assumption that \emph{all} energy gets converted into gradient energy. Our results show that the energy budget of the transition is altered and thus the signal should be modified. In~\cite{Lewicki:2022pdb} it was shown, using simplified lattice simulations for very strong transitions, that spectra sourced by bubble walls and by thin fluid shells following them have the same spectral shape. If these results hold for the scenario with significant particle production, the GW spectrum would not be significantly modified, but this issue deserves a more detailed study.
In addition to potentially modifying 
the standard GW spectrum from bubble collisions, our findings also suggest a new high-frequency GW signal, analogous to the signal formed in 
resonant preheating after inflation.
In those scenarios, where the background is homogeneous in space, the GW spectrum peaks around the maximally produced mode $k_*$, redshifted until today \cite{Giblin:2014gra, Madge:2021abk}. 
We can infer the shape of the Minkowski energy spectrum from the energy density spectrum in Fig.~\ref{fig:spectrum}. 
Due to the tachyonic mass, we can not transform the spectrum directly to Minkowski. However, if we make the simplifying assumption of massless modes in the transformation, we see that the peak of the spectrum increases in time and shifts to larger Minkowski momenta. 
Therefore, we expect that the spectrum will be spread out with respect to the usual spectrum from reheating, with a peak at frequencies much above the peak from bubble wall collisions.
We leave a precise prediction of the GW spectrum to future work, but note that the amplitude of the spectrum may be sizeable, as the amount of energy released in the phase transition is very large, and a significant fraction gets transferred to modes. 

As the total energy of the system is conserved, the growth of the fluctuations comes at the cost of the background energy, both the gradient and the kinetic components. Therefore, one can expect that the expansion of the bubble will be slower than in the case with no particle production.
Moreover, the abundantly produced particles exert friction on the bubble wall.
Even though particles produced at early stages will stay \emph{inside} the bubble, some particles are produced in front of the ``inner'' parts of the wall.
Therefore, the wall velocity, which affects the shape of the GW spectrum, has to be revisited. 
This potential slowdown may also have implications for mechanisms that rely on supercooled phase transitions for the production of heavy dark matter \cite{Azatov:2020ufh, Azatov:2021ifm} or baryogenesis \cite{Azatov:2021irb, Cataldi:2025nac, Azatov:2022tii, Baldes:2021vyz}.

Finally, we have shown that the produced fluctuations do not modify the EoS parameter compared to the background solution. As a significant fraction of energy density resides in relativistic particles long before percolation, less background energy remains to be redistributed after the phase transition. This means that the matter-domination period studied in \cite{Mansour:2026sdx, Gonstal:2025qky} should be even more difficult to achieve.

Summing up, we have shown that particles can be abundantly produced at the early stages of supercooled cosmological phase transitions. This previously overlooked phenomenon has a variety of phenomenological consequences. It modifies the existing predictions for the stochastic GW background observable with LISA and induces new (GW) signatures which warrant a dedicated study.

\section{Acknowledgements}
We would like to thank Richard Easther, Dra\v{z}en Glavan, Yann Gouttenoire, Marek Lewicki, Henda Mansour, Arttu Rajantie, Daniel Schmitt and Mateusz Zych for insightful discussions.
B\'S and  MK are supported by the National Science Centre, Poland, through the OPUS grant
no.~{\tt 2023/49/B/ST2/02782}.

\section{Data availability statement}
For the purpose of Open Access, the authors have applied
a CC-BY public copyright licence to any Author Accepted Manuscript
version arising from this submission.
The data used to prepare
Figs.~\ref{fig:evolution}--\ref{fig:spectrum} presented in this article are
publicly available at~\cite{BKIFIQ_2026}.

\section{AI statement}
The authors used OpenAI's Codex and Anthropic's Claude Code for the development of the {\tt Python} numerical code for solving the background and fluctuations equations of motion based on the previous script in {\tt Mathematica} developed solely by the authors. All calculations, numerical results, and conclusions were independently verified by the authors, who take full responsibility for the content of this work.

\bibliographystyle{JHEP}
\bibliography{bibliography}

\myendmatterrule


\pdfbookmark[1]{End Matter}{bookmark:endmatter}

\begin{center}
    \textbf{End Matter}
\end{center}

\section{Effective potential}
The zero-temperature effective potential is given by (in the Landau gauge and \MSbar{} renormalisation scheme)  
\begin{equation}
    \hspace{-0.5em}
    V_{T=0}(\varphi) = \frac 1 4 \lambda_\varphi \varphi^4 + \frac{9 M_X(\varphi)^4}{64\pi^2}\left(\log \frac{M_X(\varphi)^2}{\mu^2} -\frac{5}{6} \right),
\end{equation}
with $M_X = g_X \varphi/2$ the tree-level mass of the dark gauge field $X$ and $\mu$ the RG-scale.
The gauge-field contribution to the one-loop thermal effective potential reads
\begin{equation}
    V_{T}(\field, T)=\frac{9 T^4}{2\pi^2} J_{b}\left(\frac{M_X(\field)^2}{T^2}\right),
\end{equation}
with $J_{b}(y^2) =  \int_0^\infty \d x \, x^2 \log(1- e^{-\sqrt{x^2 + y^2}})$. 
The Arnold-Espinosa correction is given by~\cite{Arnold:1992rz, Prokopec:2018tnq, Kierkla:2022odc}
\begin{equation}
    V_{\textrm{daisy}}(\field,T)=-\frac{T}{4\pi} \left(M_{X,\textrm{th}}^3(\field,T)-M_X^3(\field)\right),
\end{equation}
with $M_{X,\textrm{th}}^2(\field,T)=M_X^2(\field)+\frac 5 {12} g_X^2 T^2.$
The relevant $\beta$-functions needed for implementing the RG-improved potential, following the procedure of~\cite{Kierkla:2022odc, Kierkla:2023von}, can be found in \cite{Chataignier:2018kay}.

\section{Mode decomposition}
The eigenfunctions of the hyperbolic Laplacian in three-dimensional space satisfy $\nabla^2_{\mathbb{H}^3} Y_{plm}(\chi,\Omega) = -(1+p^2)Y_{plm}(\chi,\Omega)$, with
\begin{align}
	Y_{plm}(\chi,\Omega) & = f_{pl}(\chi) Y_{lm}(\Omega),
    \label{eq:Y1}
    \\
    f_{pl}(\chi) & = \frac{\Gamma(ip+l+1)}{\Gamma(ip)} \frac{1}{\sqrt{\sinh \chi}} P^{-l-1/2}_{ip-1/2}(\cosh \chi),
     \label{eq:Y2}
\end{align}
with $P^{-l-1/2}_{ip-1/2}$ the Legendre function of the first kind and $Y_{lm}$ the spherical harmonic. 
The $Y_{plm}$'s form a complete orthonormal set of square-integrable functions on the unit hyperboloid.
Moreover, they obey the following addition formulas \cite{Sasaki:1994yt},
\begin{align}
\sum\limits_{lm} |Y_{plm}(\chi,\Omega)|^2 & = \f{p^2}{2\pi^2},
\\
\sum\limits_{lm} |\partial_\chi Y_{plm}(\chi,\Omega)|^2 & = \f{p^2(p^2+1)}{6\pi^2},
\\
\f{1}{\sinh^2\chi}\sum\limits_{lm} |\nabla_{\mathbb{S}^2}Y_{plm}(\chi,\Omega)|^2 & = \f{p^2(p^2+1)}{3\pi^2}.
\end{align}

\section{Renormalization}
The $\braket{\delta \hat \varphi^2}$ term in Eq.~\eqref{eq:EOM_background_no_decomposition}, $\tilde\rho$ in Eq.~\eqref{eq:milne_energy_density}, and $\tilde{P}$ in Eq. \eqref{eq:milne_pressure} are UV-divergent, and need to be renormalized. 
Since UV modes are completely unaffected by the presence of the bubble, these divergences are identical to the usual divergences in \mbox{$\phi^4$-theory}, and can be subtracted with the usual counterterms (cosmological constant, mass, and coupling).
This is not immediately obvious from Eqs.~\eqref{eq:energydens} and \eqref{eq:press}, which seem to feature $\chi$-dependent divergences.
For large momenta, away from the tachyonic regime, $(p/s)^2 \gg |V''(\varphi(s))|$, the adiabatic expansion holds~\cite{Herring:2024nqg}, and the divergences are captured by the zeroth-order adiabatic contribution. 
It is given by the vacuum solutions of
Eq.~\eqref{eq:EOM_fluctuations} 
-- the Hankel functions.
By using the integral representation of the Hankel functions, it can be shown that the $\chi$-dependence completely cancels out of the divergence of the energy density, and that it is identical to the result obtained with Minkowski plane-wave vacuum solutions. 

The correspondence of the divergences in Minkowski and Milne coordinates allows us to do the following. 
Suppose we want to evaluate the divergent integral $\int^\infty_0 p^2 \mathcal G_p$, with $\mathcal G_p$ some combination of mode solutions. 
We can add and subtract $\int^\infty_0 p^2 \mathcal G_{p,\rm{vac}}$, where $\mathcal{G}_{p,\rm{vac}}$ is evaluated using the asymptotic vacuum modes. In the subtracted part, we use the Milne vacuum, whereas in the added part, the Minkowski one.
We know that the divergence of the Minkowski term is cancelled by the usual counterterms, so we only need to consider its finite part.
So we are left with $\int^\infty_0 p^2 (\mathcal G_p -\mathcal G_{p,\rm{vac}})$. 
To evaluate this numerically, we choose a scale $\Lambda$, such that $\int^\infty_{\Lambda} p^2(\mathcal G_p - \mathcal G_{p, \rm{vac}}) \rightarrow 0$ up to a chosen accuracy. 
We then only need to evaluate the finite integral $\int_0^\Lambda p^2 (\mathcal G_p -\mathcal G_{p,\rm{vac}})$ and the finite Minkowski part numerically.

In practice, in Eq.~\eqref{eq:EOM_background_no_decomposition} we subtract the term with  $\braket{\delta\hat\varphi^2}$ evaluated with the high-$p$ asymptotics of the instantaneous vacuum modes, $\delta\varphi_{p,\text{vac}} = \f{\sqrt{\pi}}{2s} e^{\pi p/2} H_{ip}^{(2)}(m s)$, where $H^{(2)}$ is the Hankel function of the second kind with the instantaneous mass, $m^2=V''(\varphi(s))$, and add the same term evaluated with Minkowski vacuum. The Minkowski term reproduces the scalar correction to the effective potential, which is renormalised by the potential counterterms. This is performed only for the non-tachyonic modes~\cite{Herring:2024nqg} by adding a factor of $\Theta \left(p/s-k_{\text{max}}\right)$, where $k_{\text{max}}$ represents the highest momentum scale at which the field is tachyonic $k_{\text{max}} = \max(\sqrt{-V''(\varphi)})$. 
This gives the EoM as
\begin{align}
    &\ddot{\varphi} + \f3s\dot{\varphi} + V_{\text{eff}}'(\varphi) + V'''(\varphi)  \times \int\limits_0^\infty \f{p^2\d p}{4\pi^2} \Big[ 
    \\ 
    &|\delta\varphi_p|^2 - 
    \Theta \left(\f ps-k_{\text{max}} \right) \f{1}{2 s^2\sqrt{p^2 + (m s)^2}} \Big] = 0, \nonumber
\end{align}
where
\begin{align}
    V_{\text{eff}}(\varphi) & = V(\varphi) + V_{\text{CW}}(\varphi) - (m^2)^2 \mathcal{F}\left(\f{k_{\text{max}}}{|m^2|^{1/2}}\right), 
    \\
    V_{\text{CW}}(\varphi) & = \f{(m^2)^2}{64\pi^2}\left[ \log \left(\f{m^2}{\mu^2} \right)  -\f32\right], 
\end{align}
is the renormalised effective potential including the effects of the tachyonic modes and $\mathcal{F}$ given by Eq. (V.10) in \cite{Herring:2024nqg}. 
This effective potential must also be taken into account when calculating the (potential) energy density of the background field.
Note, however, that in practice, the contributions of $V_{\rm CW}$
and $\mathcal F$ are negligible. The reason is that they do not depend on the tachyonically enhanced modes and are proportional to $\lambda_\varphi^2$, which scales as $\sim g_X^8$.

To renormalize $\tilde{\rho}$ and $\tilde{P}$, we introduce the auxiliary quantities of Milne vacuum energy density subtraction and vacuum pressure subtraction:
\begin{align}
    \tilde{\rho}_{\text{sub}}  = \int\limits_0^\infty \f{p^2 \d p}{4\pi^2} \Theta \Big(\f {p}{s}- & k_{\text{max}} \Big) \f{p^2 + 1 + (ms)^2}{s^4 \sqrt{p^2 + (ms)^2}}
    \\
    & \times \left(1 + \f{(ms)^2}{2(p^2 + (ms)^2)^2} \right),
    \nonumber
    \\
    \tilde{P}_{\text{sub}} = \int\limits_0^\infty \f{p^2 \d p}{4\pi^2} \Theta \Big(\f ps - & k_{\text{max}} \Big) \f{p^2 + 1}{3s^4 \sqrt{p^2 + (ms)^2}}
    \\
    & \times \left(1 + \f{(ms)^2}{2(p^2 + (ms)^2)^2} \right), 
    \nonumber
\end{align}
which are the asymptotic expressions for the Milne energy density \eqref{eq:milne_energy_density} and pressure \eqref{eq:milne_pressure} for the instantenous-mass vacuum solution $\delta\varphi_{p, \text{vac}}$, agreeing up to the logarithmic divergence. We subtract these quantities from the fluctuations' Milne energy density and pressure and add the corresponding $V_{\text{eff}}$ contribution to the background energy density and pressure.

\section{Numerical implementation} 
We initialise the field background and mode values at small $s_{\text{min}} = 10^{-5} \text{ GeV}^{-1}$ (instead of the numerically unavailable $s\to0$).
We calculate the $\braket{\delta\hat\varphi^2}$ momentum integral as the trapezoidal Riemann sum over the lattice momenta $p_i$.
For better numerical cancellation, the vacuum-subtraction integrals are also evaluated as the trapezoidal Riemann sums of their respective integrands.
As both energy density (calculated from the mode solutions) and energy density vacuum-subtraction diverge as $s \to 0$ (with their difference remaining finite), their numerical calculation loses precision for low $s$. Therefore, we include an extra $\Theta(s - s_{\text{low}})$ in the calculation of the bubble's energy budget, with an appropriately chosen $s_{\text{low}} \sim O(10^{-2}) \text{ GeV}^{-1}$, high enough such that the numerical singularity disappears, but low enough such that important production of fluctuations is included in the calculation. We have verified the robustness of the solution against moderate changes in the choice of $s_{\rm low}$. We also used a numerical padding prefactor in the implementation of the low-momentum tachyonic integral cutoff $\Theta(p/s-k_{\text{max}}) \to \Theta(p/s-(1+\varepsilon)k_{\text{max}})$, with $\varepsilon = 0.05$ and accordingly used $(1+\varepsilon)k_{\text{max}}$ in the argument of $\mathcal{F}$.

\vspace{\stretch{1}}

\end{document}